\documentclass[twocolumn,showpacs,amsmath,amssymb]{revtex4}
\usepackage{graphicx}

\begin{document}

\title{``Nodal gap'' induced by the incommensurate diagonal spin density  modulation in underdoped high-$T_c$ superconductors}

\author
{Tao Zhou$^{1}$, Yi Gao$^{2}$, and Jian-Xin Zhu$^{3}$}

\affiliation{$^{1}$College of Science, Nanjing University of Aeronautics and Astronautics, Nanjing 210016, China\\
$^{2}$Department of Physics and Institute of Theoretical Physics,
Nanjing Normal University, Nanjing 210023, China\\
$^3$Theoretical Division and Center for Integrated Nanotechnologies, Los Alamos National Laboratory, New Mexico 87545, USA
}

\date{\today}
\begin{abstract}
Recently it was revealed that the whole Fermi surface is fully gapped for several families of underdoped cuprates.
 The existence of the finite energy gap along the $d$-wave nodal lines (``nodal gap'') contrasts the common understanding of the $d$-wave pairing symmetry, which challenges the present theories for the high-$T_c$ superconductors. Here we propose that the incommensurate diagonal spin-density-wave order can account for the above experimental observation. The Fermi surface and the local density of states are also studied. Our results are in good agreement with many important experiments in high-$T_c$ superconductors.
\end{abstract}
\pacs{74.25.Jb, 74.72.Kf, 74.20.Rp}
 \maketitle

The energy gap is one of the most important properties in the studies of the high-$T_c$ superconductors.
Recently,  measurements of the energy gaps by angle-resolved photoemission
spectroscopy (ARPES) in lightly doped high-$T_c$ materials revealed the existence of the non-zero energy gap along the diagonal directions of the Brillouin zone (also referred to the ``nodal gap'')~\cite{har,vis,val,raz,pen,dra}.
The earliest indication of fully gapped single-particle excitation was reported in Ref.~\cite{she}.
The existence of the nodal gap seems to be generic. It has been observed in several families of cuprates~\cite{har,vis,val,raz,pen,dra,she}.
Moreover, it has been reported that the nodal gap exists in the antiferromagnetic (AF) state~\cite{pen,dra}, the spin glass region~\cite{val}, as well as the superconducting and normal states for deeply underdoped region~\cite{har,vis,raz,she}.
This result is surprising and
contrasts the usual understanding of the $d$-wave superconducting pairing or the conventional pseudogap behavior, both should generate energy nodes along diagonal lines of the Brillouin zone.
The opening of the nodal gap in the AF state is also intriguing. It was reported that the commensurate AF order forms at $140$ K, well above the temperature that the nodal gap opens, which is only $45$ K~\cite{dra}.
 The above experimental observations challenge the present theory for high-$T_c$ superconductors. A theoretical elucidation of the nodal gap is highly demanded.

The theoretical understanding of the gap behavior and its relation with the superconductivity are of fundamental interest and may be essential to understand the superconductivity.
As is known, even in the superconducting state the quasiparticle energy gap is not necessarily tied to the superconducting order parameter.
It is rather important to explore the physics behind the gap-like feature.
Very recently the origin of the nodal gap has attracted tremendous attention. Many groups have attempted to give possible theoretical scenarios for this issue. Possible explanations include Coulomb disorder effects~\cite{chen}, fluctuating competing AF order~\cite{atk}, topological superconductor~\cite{yuan}, the coexistence of AF and superconducting order~\cite{das}, and the coexistence of $d_{x^2-y^2}+id_{xy}$-wave superconducting order and $d_{xy}$-wave AF order~\cite{gup}.
So far this issue is still far from obvious and no consensus has been reached.

Motivated by the above experimental observations and theoretical attempts and in order to give a more definitive explanation for the nodal gap,
in this Letter,  we start from a phenomenological model in the presence of an incommensurate diagonal  spin-density-wave (ID-SDW) order with the wave vectors ${\bf Q}=(\pi\pm\delta,\pi\pm\delta)$
 to elaborate its effect on the
spectral function. The major advances made in the present work includes: i) The presence of the ID-SDW order
has been revealed by neutron scattering experiments and it is understandable within the Fermi surface nesting picture; ii) The nodal gap is robust in presence of the above ID-SDW order. A rather reasonable explanation can be presented; iii) The calculated renormalized  Fermi surface in the normal state is in good agreement with previous experiment;  iv) The modulated real space local density of states and its Fourier transformation are also investigated and it is qualitatively consistent with previous scanning tunneling microscopy (STM) experiments.

Our starting phenomenological model includes the superconducting term and an ID-SDW order term, which expressed as,
\begin{equation}
H=H_{SC}+H_{S}\;,
\end{equation}
where the superconducting Hamiltonian is expressed as,
\begin{eqnarray}
H_{SC}=-\sum_{{\bf ij},\sigma}t_{\bf ij}c^{\dagger}_{{\bf i}\sigma}c_{{\bf j}\sigma}-\mu\sum_{{\bf i}\sigma}c^{\dagger}_{{\bf i}\sigma}c_{{\bf i}\sigma}\nonumber \\+\sum_{\bf ij}(\Delta_{\bf ij}c^{\dagger}_{{\bf i}\uparrow}c^{\dagger}_{{\bf j}\downarrow}+h.c.)\;.
\end{eqnarray}
In the present work, we assume phenomenologically the SDW ordered  periodically, expressed as
\begin{equation}
H_S=\sum_{\bf i Q_s}V_s S^z_{\bf i} e^{i{\bf R_i}\cdot {\bf Q_s}}.
\end{equation}

The above Hamiltonian can be transformed to the momentum space by taking into account the $d_{x^2-y^2}$-wave superconducting pairing, the nearest-neighbor and next-nearest-neighbor hopping, which is rewritten as,
\begin{equation}
H_{SC}=\sum_{{\bf k}\sigma}\varepsilon_{\bf k}c^{\dagger}_{{\bf k}\sigma}c_{{\bf k}\sigma}+\sum_{\bf k}(\Delta_{\bf k}c^{\dagger}_{{\bf k}\uparrow}c^{\dagger}_{-{\bf k}\downarrow}+h.c.),
\end{equation}
where $\varepsilon_{\bf k}=-2t(\cos k_x+\cos k_y)-4t^\prime \cos k_x\cos k_y-\mu$, and $\Delta_{\bf k}=\Delta_0 (\cos k_x-\cos k_y)$,
and $H_S$ is expressed as
\begin{equation}
H_{S}=\sum_{{\bf k}\sigma {\bf Q_s}}({V}\sigma c^{\dagger}_{\bf k\sigma}c_{{\bf k}+{\bf Q_s},\sigma}+h.c.),
\end{equation}
with $V=V_s/2$ is the ID-SDW order magnitude.

The wave vectors of the ID-SDW order (${\bf Q_s}$) can be determined qualitatively from the neutron scattering experiments.
 The static diagonal incommensurate order has been observed in the deeply underdoped region~\cite{wak,waki,mat,wakim,fuj,eno}.
 As the doping density increases, the spin modulation along the parallel direction with the wave vectors $(\pi,\pi\pm\delta)$ and $(\pi\pm\delta,\pi)$ was observed, accompanied by the appearance of the superconductivity.
It was revealed that the diagonal spin order could persist into the
superconducting state and coexist with the parallel spin order~\cite{fuj}.
Very recently, based on the muon spin rotation measurement, the ID-SDW order was also observed in the AF state, with the temperature about 30K. It is
much lower than the AF Neel temperature, while close to that of the nodal gap being observed~\cite{dra}.
Therefore,  the ID-SDW order and the nodal gap appear in the same region  experimentally. According to the above experimental results, in the present work,
we consider the four  incommensurate scattering wave vectors ${\bf Q_{s}}=(\pi\pm\delta,\pi\pm\delta)$, with $\delta=0.15\pi$ ($\delta$ is obtained from the Fermi surface nesting vector, as discussed below).
Hereafter, if not specified otherwise,
the parameters are used as $t=1$, $t^\prime=-0.3$, $\mu=-0.857$ (corresponding to the doping $x=0.08$), $\Delta_0=0.25$, $V=0.1$. We have checked numerically that our results
are not sensitive to the reasonable changes of the chosen parameters.

\begin{figure}
\centering
  \includegraphics[width=3in]{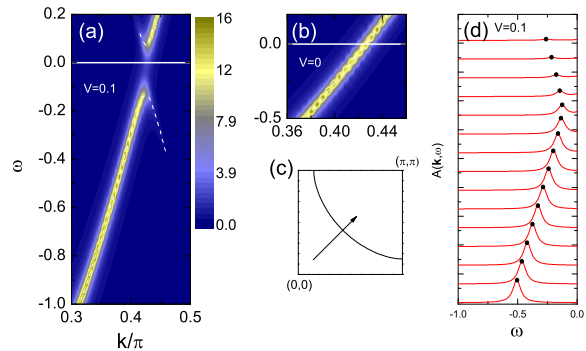}
\caption{(Color online) (a-b) Intensity plots of the spectral function with and without the ID-SDW order, respectively. (c) The normal state Fermi surface. The arrow indicates the cut along which panels (a,b,d) were taken. (d) The energy dependence of the spectral function along the arrow in (c) (from bottom to top).   }
\end{figure}

For the incommensurability under consideration $(\delta=0.15\pi=2\pi*3/40)$, the whole Brillouin zone is divided into $40\times 40=1600$ parts. The Hamiltonian with the superconducting pairing can be written as $3200\times 3200$ matrix. Then the retarded  green's function $G({\bf k},\omega+i\Gamma)$ can be obtained
through diagonalizing the Hamiltonian. The quasiparticle spectral function $A({\bf k},\omega)$ is given from the retarded Green's function with $A({\bf k},\omega)=-\text{Im}G({\bf k},\omega+i\Gamma)/\pi$.

We show in Figs.~1(a) and 1(b) the spectral function along the diagonal direction (along the cut indicated in Fig.1(c)) with and without the ID-SDW order, respectively. The dashed lines are the quasiparticle dispersions.
The spectral functions as a function of the energy (EDCs) along the diagonal direction are plotted in Fig.~1(d).
As is seen, the quasiparticle energy decreases as the wave vector moves towards the Fermi surface.
An obvious gap exists in the presence of the ID-SDW order, as shown in Fig.~1(a). We can also see clearly that the gap closes and the quasiparticle dispersion crosses the Fermi momentum $K_F$ for $V=0$, as is seen in Fig.~1(b). We also checked numerically that the above results are in fact independent on the $d$-wave pairing magnitude $\Delta_0$ and the nodal gap exists when we set $\Delta_0=0$ (not shown here). Therefore, the above nodal gap should exist both in the superconducting state and the normal state. Our results for the nodal gap are qualitatively consistent with the experiments~\cite{har,vis,val,raz,pen,dra,she}.

The momentum dependence of the energy gap along the Fermi surface is studied in Fig.~2. The EDCs with different Fermi surface angle $\theta$ [defined in Fig.~2(b)] are plotted in Fig.~2(a). We define the energy gaps as the peak positions of EDCs. Then the energy gap as a function of the Fermi angle is shown in Fig.~2(c). As is seen, the energy gap is significantly anisotropic. It reaches the maxima value at the Brillouin boundary and decreases when the wave vector moves towards the diagonal direction. It reaches the minimum value at the diagonal direction. The $d$-wave gap magnitude is also plotted in Fig.~2(c) for comparison. The observed energy gap and the $d$-wave gap are nearly the same near the antinodal direction. Near the diagonal direction
 the gap is different from the $d$-wave one, namely,
 an obvious finite gap exists due to the presence of the ID-SDW order.
The above results are qualitatively consistent with the experimental observations in the superconducting state~\cite{vis}. We also note that very recent ARPES experiment on the insulating samples has also revealed that the energy gap is anisotropic, i.e., it reaches the maxima value at the
Brillouin zone boundary and is minimum at the diagonal direction~\cite{pen}.
This experimental result can also be explained qualitatively based on our model.
Note that the origin of the energy gap in deeply underdoped high-Tc materials is indeed complicated. There may exist several candidate competing orders.
We propose that the energy gap near the $d$-wave nodal points is still due to the ID-SDW order.
The gap near the Brillouin boundary is generated by another order (e.g., the $d$-density wave order~\cite{chak} or the AF order~\cite{note}).
The coexistence of two orders in the insulating sample are also supported by very recent experiments~\cite{dra}.
Theoretically, the energy gap produced by the $d$-density-wave order or AF order should be maximum near the hot spots (the crossing points between the normal state Fermi surface and the magnetic Brillouin zone). For hole doped samples, the hot spots are close to the antinodal points. Thus the anisotropic gap in non-superconducting materials~\cite{pen} is understandable. We have also checked numerically (not shown here) that similar anisotropic behavior can be reproduced  with the model including both the ID-SDW order and the $d$-density-wave order (or the AF order).

\begin{figure}
\centering
  \includegraphics[width=3in]{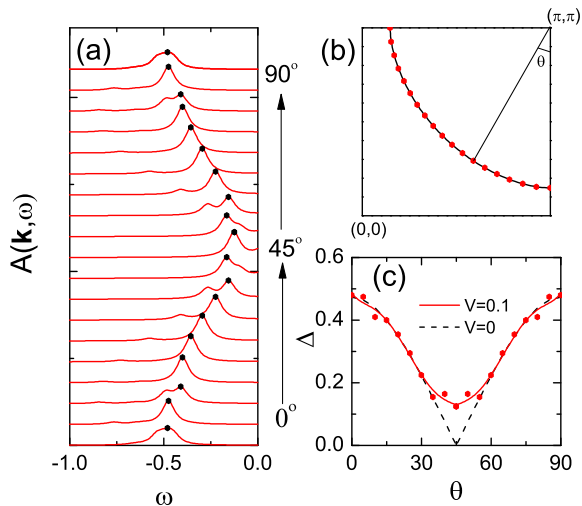}
\caption{(Color online) (a) The spectral function as a function of the energy with different Fermi angle $\theta$, with the points and the Fermi angle are shown in (b).
(c) The energy gap as a function of the Fermi angle. The closed circles are the energy gap obtained from panel (a). The red solid line is a polynomial fitting for the data. The dashed line is the $d$-wave superconducting gap magnitude. }
\end{figure}

The explanation of the ID-SDW order can be given based on the Fermi surface nesting picture.  The normal state Fermi surface is shown in Fig.~3.
As is seen, the tangent lines of the Fermi surface curve are parallel at the $d$-wave nodal points, revealing the Fermi surface nesting feature.
The corresponding nesting vector is marked in Fig.~3, for which the incommensurability can be obtained from the Fermi momentum along the diagonal direction.
The ID-SDW order comes mainly from such node-to-node excitations. As such,  the existence of the nodal gap can be immediately understood. Namely, the vector of the ID-SDW order connects different nodal points of Fermi surface. The electron hopping between these points can occur due to the ID-SDW order. This destroys the state of the quasiparticle near nodal points and an energy gap opens.

\begin{figure}
\centering
  \includegraphics[width=3in]{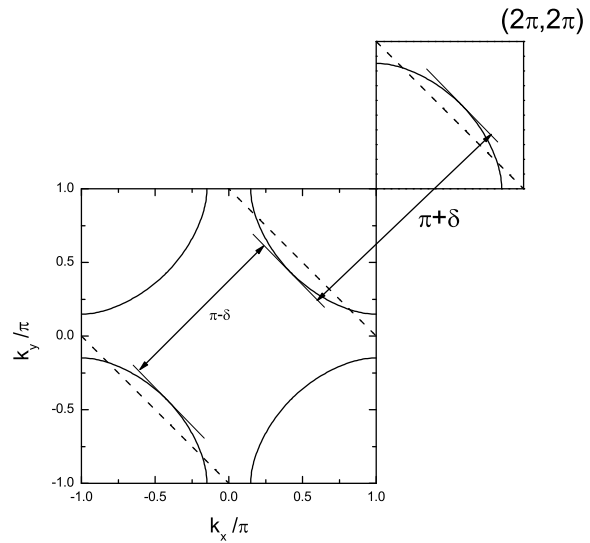}
\caption{The normal state Fermi surface ($\varepsilon_{\bf k}=0$). The diagonal nesting wave vectors are indicated.  }
\end{figure}

The signatures of this ID-SDW order can be probed by studying the normal state Fermi surface. Previously many interesting results for the Fermi surface of underdoped cuprates have been obtained by ARPES experiments. One important observation is that the Fermi surface is gapped near the antinodal direction and leaves an ungapped Fermi arc~\cite{nor}. This conventional pseudogap behavior is still unsolved and not concerned in the present work. On the other hand, the electronic structure along diagonal direction is also non-trivial. It was revealed that the spectral weight is low and the quasiparticle peak is broad near the nodal direction for underdoped La$_{2-x}$Sr$_x$CuO$_4$ samples~\cite{ino,inoa}. Another interesting result is the observation of the Fermi pocket in the underdoped samples~\cite{men}. It was revealed that the Fermi pocket coexists with the Fermi arc, and exists only in the underdoped samples. Interestingly, the experimentally observed Fermi pocket is not symmetrical with respect to the $(0,\pi)$ to $(\pi,0)$ line. Thus the $d$-density-wave order or AF order may not account for the Fermi pocket. It was also proposed in Ref.~\cite{men} that the incommensurate diagonal density-wave may potentially explain their results.

\begin{figure}
\centering
  \includegraphics[width=3in]{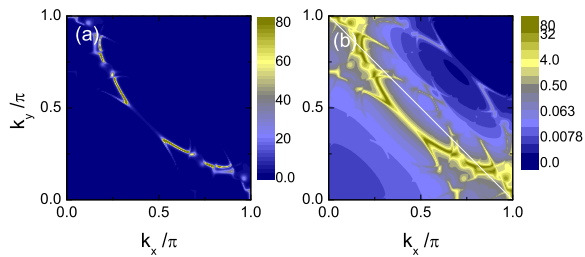}
\caption{(Color on line) (a) The intensity plot of the zero energy spectral function $A({\bf k},\omega=0)$ with $\Delta_0=0$ and $V=0.1$.  (b) The same with (a)
while the logarithmic scale is used. }
\end{figure}

The numerical results of the normal state zero energy spectral function $[A({\bf k},\omega=0)]$ ($\Delta_0=0$, $V=0.1$) is presented in Fig.~4. The normal state Fermi surface can be obtained through the peaks of the spectral function. As is seen in Fig.~4(a), the spectral weight near the diagonal direction is quite low, consistent with the experimental observations~\cite{ino,inoa}. This is due to the node-to-node scattering caused by the ID-SDW.  When the spectral function is plotted in a logarithmic scale,  the weak features of the spectral function is revealed more clearly.
 As is seen, besides the main Fermi surface, contributed by the normal state energy band $\varepsilon_{\bf k}$, another band with much lower spectral weight, can be seen clearly. Then a Fermi pocket forms. Here the Fermi pocket is non-symmetrical and not centered at ($\pi/2,\pi/2$). The above results are qualitatively consistent with the recent experimental observation~\cite{men}.

\begin{figure}
\centering
  \includegraphics[width=3in]{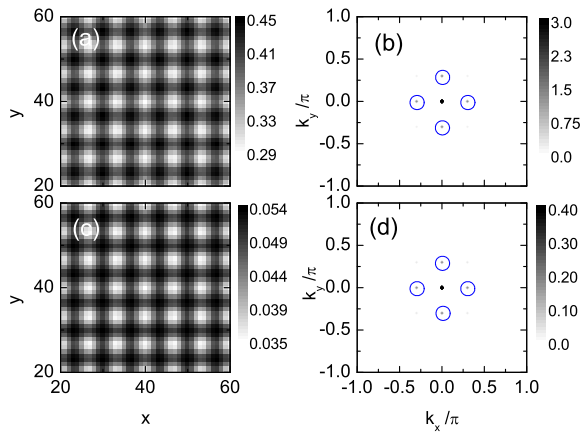}
\caption{(Color online) The LDOS [$\rho_{\bf i}(\omega)$] and FT-LDOS [$Z({\bf q},\omega)$] in the normal state and superconducting state with $V=0.1$ and $\omega=0.1$. (a) LDOS in the normal state with $\Delta=0$. (b) FT-LDOS in the normal state. (c) LDOS in the superconducting state with $\Delta_0=0.25$. (d). FT-LDOS in the superconducting state.   }
\end{figure}

The relationship of the SDW and charge-density-wave (CDW) orders has been an important point and attracted intensive attention previously.
Experimentally the charge order could be detected through the STM experiments.
One prominent feature is
 the ``checkerboard structure'' from the energy-dependent local density of
states (LDOS). It was first reported to exist in vortex cores of optimally doped materials, with a two-dimensional modulation along Cu-O bond directions~\cite{hof,jxz1,jxz2}.
Later experiments also observed similar modulation in the superconducting samples without the magnetic field~\cite{how,han}. In the meantime, the charge order can also be revealed in more detail through the Fourier transform of the
LDOS (FT-LDOS). The non-dispersive peaks would be observed in the presence of the charge modulation.
The periodicity can be determined through the peak positions in the momentum space.
An incommensurate charge modulation with the periodicity of about $4.5a\sim 4.7a$ was reported in the normal and superconducting states~\cite{ver,fan,mce}.

We now study numerically the real space modulation induced by the ID-SDW order. Diagonalizing the Hamiltonian [Eq.~(1)], we can obtain the LDOS $\rho({\bf i},\omega)$ numerically. To compare with the STM experiments, we also define the Fourier transformation of the LDOS (FT-LDOS), which is expressed as $Z({\bf q},\omega)=\sum_{\bf i}\rho_{\bf i}(\omega)\exp(i{\bf R_i}\cdot {\bf q})$.
The numerical results for the LDOS and FT-LDOS are presented in Fig.~5. The LDOS in the normal state with the energy $\omega=0.1$ is plotted in Fig.~5(a). The checkerboard parttern is revealed clearly. Fig.~5(b) is the FT-LDOS spectra. There exist four peaks at the wave vector $(0,\pm 0.3\pi)$ and $(\pm0.3\pi,0)$ (indicated with circles). The above results are robust and qualitatively the same for different energies. The LDOS and FT-LDOS in the superconducting state are plotted in Figs.~5(c) and 5(d). As is seen, the intensities decrease due to the existence of the superconducting gap. While the main results are qualitatively the same with those of the normal state. Interestingly, although here we consider the ID-SDW order in the starting model, the modulations of LDOS are along the CuO bond directions. This is consistent with the STM observations. It is also worthwhile to point out that the relation between the SDW order and charge order is still an open question.
If both have the same origins, then a simple relation for the incommensurability of the SDW order $\delta_s$ and the CDW order $\delta_c$
should satisfies: $\delta_c=2\delta_s$~\cite{tra}. Experimentally this relation is consistent with the observations in La$_{2-x}$Sr${_x}$CuO${_4}$~\cite{tra,cro} and La$_{2-x}$Ba${_x}$CuO${_4}$~\cite{huc} samples. However, it was also revealed that this relation is not satisfied in the Bi$_{2}$Sr$_{2-z}$La${_z}$CuO$_{6+x}$~\cite{wise,com} and YBa$_2$Cu$_3$O$_y$ samples~\cite{bla}.

In summary, based on a phenomenological model, we elaborate that an ID-SDW order can cause a finite gap along the $d$-wave nodal line. This is in good agreement with recent experimental observations. The origin of the ID-SDW order can be explained through the Fermi surface nesting picture. The normal state Fermi surface and the local density of states are also studied. The results are qualitatively consistent with the experiments.

This work was supported
by the NSFC (Grant No. 11374005 and No. 11204138), the NCET (Grant No. NCET-12-0626), NSF of Jiangsu Province of China (Grant No. BK2012450), Jiangsu Qingnan engineering project, and U.S. DOE  Office of Basic Energy Sciences.

\end{document}